\documentclass{basi}
\usepackage[british]{babel}
\usepackage[varg]{txfonts}
\begin{document}

\title[Spectral Line Changes with Solar Transients]{Spectral Line Profile Changes Associated with Energetic
Solar Transients} \author[A.\ Ambastha \& R.~A.\ Maurya]{Ashok
Ambastha\thanks{email: \texttt{ambastha@prl.res.in}} and
    Ram A.\ Maurya\thanks{Current address: Astronomy Program, Department of Physics and Astronomy,
    Seoul National University, Seoul 151-747, Korea., email: \texttt{ramaurya98@gmail.com}}\\
    Udaipur Solar Observatory, Bari Road, P.O. Box 198, Udaipur 313 001\\}

%\pubyear{2011}
%\volume{00}
%\setcounter{page}{17}

\date{Received --- ; accepted ---}

\maketitle

\label{firstpage}

\begin{abstract}
Solar energetic transients occurring in solar atmosphere are
associated with catastrophic release of energy in the solar corona.
These transients inject a part of their energy by various physical
processes to the deeper, denser photospheric layer at which velocity
and magnetic fields are measured using suitable spectral lines.
Serious questions have been raised about the nature of the observed
magnetic (and velocity) field changes associated with energetic
transients as their measurements are expected to be affected by
flare-induced line profile changes. In this paper, we shall discuss
some recent progress on our understanding of the physical processes
associated with such events.
\end{abstract}


%------------------------------------------------------------------------------%
\section{Introduction}\label{sec:intro}

Solar flares are sudden transient, localized eruptions of energy in
the solar atmosphere. These events occur over a large range in space
and time; lasting from a few seconds to several hours.
Intensification of radiation over the entire electromagnetic
spectrum, release of energetic particles and ejection of plasma
clouds are some of the observed phenomena associated with flares.
Flares produce short-lived waves traveling away from their site,
such as, Moreton waves, and blast waves. More recently,
flare-associated helioseismic effects are also reported, such as,
compact acoustic sources, seismic waves, and amplification of
acoustic or p-modes (Ambastha, Basu and Antia, 2003).

It is generally accepted that solar transients,viz.,  flares and Coronal
Mass Ejections (CMEs), derive their energy from stressed magnetic
field. Therefore, magnetic stress or non-potentiality is expected to
relax to a lower energy state after the release of excess energy
from the the solar active region (AR). A wide variety of results are
reported on the pre- and post-flare changes in magnetic field
(Ambastha et al. ,1993; Kosovichev \& Zharkova, 2001; Mathew \&
Ambastha, 2000; Wang et al., 2002; Sudol \& Harvey, 2004).
However, serious questions have also been raised about the nature of
observed changes as the measurements are expected to be affected by
flare-induced spectral line profile changes (Harvey, 1986; Qiu and
Gary, 2003). We shall discuss here some observed phenomena occurring
in the photospheric magnetic and velocity fields during large flares
and interpret them with the help of the recently available spectral
data.

\section{Observed Flare/CME related changes}\label{sec:using}

To distinguish between the normal magnetic (velocity) field evolution of solar ARs,
and the changes considered to be associated with flares and CMEs, high
sensitivity magnetic (Doppler velocity) data are required at
high-cadence and high spatial resolution. Such data are available
from the ground-based Global Oscillation Network Group (GONG) since
1995, and the space-borne Solar Oscillations and Heliospheric
Observatory (SOHO) during 1996-2009. The space-borne measurements
are considered to be better as they are not affected by atmospheric
degradations. More recently, very high quality data are being
provided by the Solar Dynamics Observatory (SDO)-Helioseismic and
Magnetic Imager (HMI) subsequent to its launch in 2010.

Both abrupt and persistent changes in observed magnetic field parameters have been
reported that last over a few minutes during the impulsive phase of
many energetic solar transient events, i.e., flares and CMEs.
However, an anomalous magnetic flux polarity reversal was first
reported during the large X-class flare of 6 April 2001 using MDI
magnetograms (Qiu and Gary, 2003). Rapidly moving transient
features, with speeds ranging from 30 to 50 km-s$^{-1}$, were
reported also in magnetic and Doppler images of AR NOAA 10486 during
the super-flares X17/4B  (2003 October 28) and X10/2B (2003 October
29) by Maurya \& Ambastha (2009: hereafter MA09). These transient
features seen in both the GONG and MDI data were also the sites of
magnetic polarity reversals during the impulsive phase of the energetic flares
(Figure 1).

\begin{figure}
\centerline{\includegraphics[width=4.0cm]{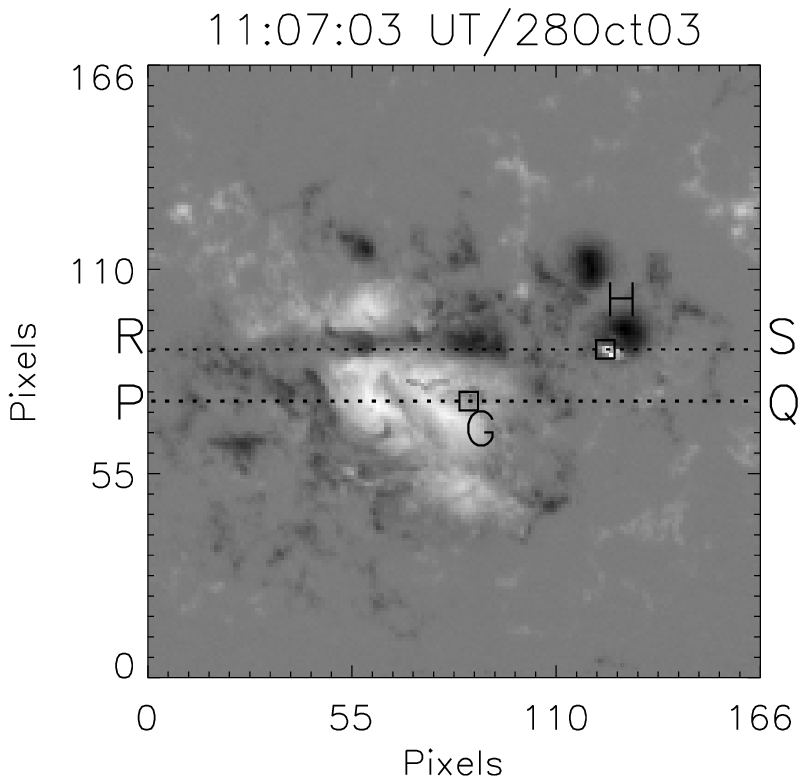} \qquad
            \includegraphics[width=4.0cm]{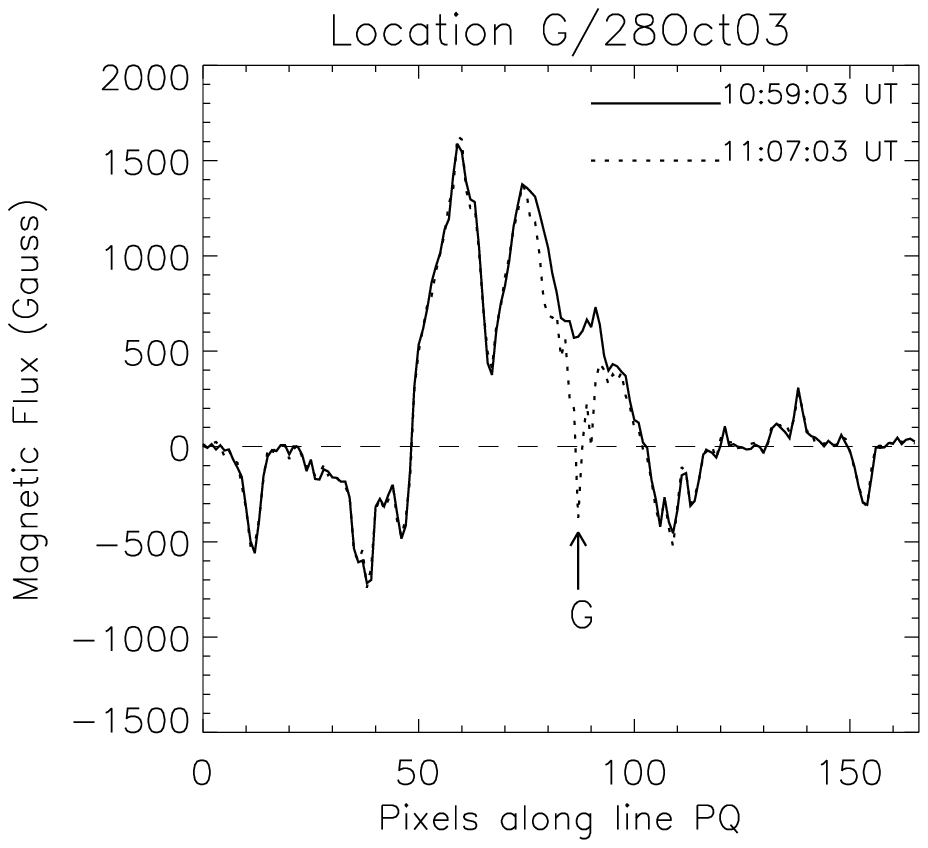} \qquad
            \includegraphics[width=4.0cm]{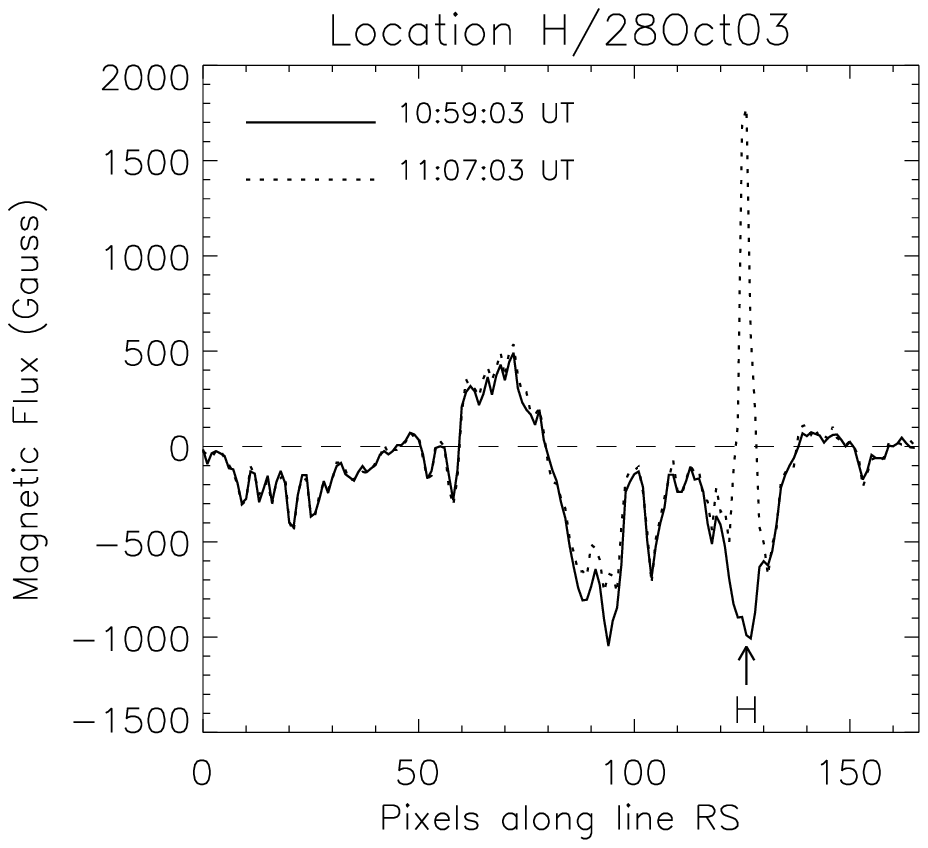} }
\caption{(a) MDI magnetogram taken around the peak phase of the 2003
October 28/11:07:03 UT.  Solid and dotted profiles show
the magnetic flux measured along lines RS (b) and PQ (c) at 10:59:03
UT and 11:07:03 UT, respectively. Polarity reversals are evident at
points G and H around 11:07:03 UT, i.e., the peak phase of the
flare. Else where along the two lines, magnetic flux remained nearly
unchanged.\label{f:one}}
\end{figure}

Such magnetic polarity sign reversals were first reported several
years ago by Zirin \& Tanaka (1981) in some large flares. They
attributed it to the heating of lower atmosphere by the flare such
that the core of the absorption line, used for measurement, turned
to emission (Patterson, 1984). However,  only a few cases of
anomalous magnetic and Doppler transients driven by the flares have
been reported in recent years. It appears that the transients may
reach detectable levels only in very energetic flares. Both the
flares in MA09 were extremely energetic white-light (WL) events.
More recently, Kosovichev (2011) have reported ``sunquake'' sources
in the regions of these features and suggested these to arise due to
thermal and hydrodynamic effects of high-energy particles heating
the lower atmosphere. One would like to ask whether these features
are related to the line profile change (Ding et al. 2002; Qiu \&
Gary, 2003) or to other physical processes such as the impact of a
shock wave (Zharkova \& Zharkov, 2007) on the line-production
region.

Due to the non-availability of the required spectral data, the
transients in the super-flares of 2003 (MA2009) were examined using
only the imaged data of the flares at various wavelengths and
inferences were drawn based on indirect methods. But the solution of
the problem lies in analyzing spectral profile shapes before, during
and after major flare events, not available thus far. Therefore, the
question of spectral line profile changes in relation to the
observed magnetic anomaly remained unresolved.

\section{Spectral line profile changes associated with flares}

Photospheric magnetic field measurements are sensitive to changes in
the line profile during the intense heating phase of solar flares
(Ding et al., 2002). They modeled the Ni {\sc i} 6768 \AA~line
profiles using radiative transfer to explain the reversals observed
during flares. However, the sign reversals as observed by MDI and
GONG are difficult to be explained by the sudden heating of the
lower atmosphere. This is because the Ni {\sc i} line (used in both
MDI and GONG) is formed in the temperature minimum region and is
rather stable against temperature changes (Bruls, 1993).

\begin{figure}
\centerline{\includegraphics[width=11cm]{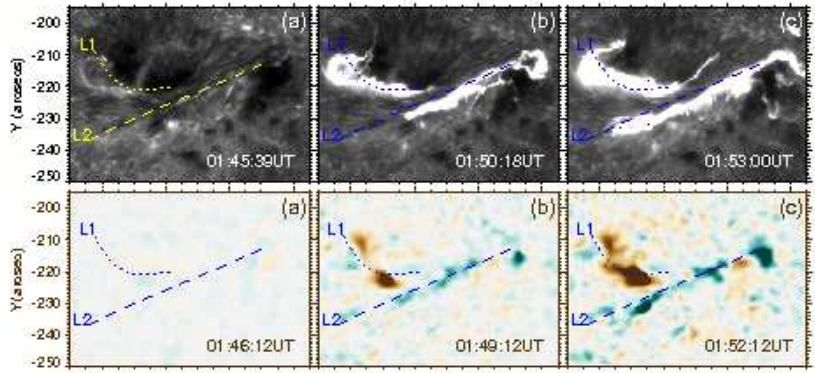}}
\caption{(Top row): Ca {\sc ii} H 3968 \AA~ filtergrams (observed by
Hinode Solar Optical telescope) showing the spatio-temporal
development of the two-ribbon X2.2 flare in AR NOAA 11158 on 2011
February 15. (Bottom row): Consecutive difference images of HMI
magnetograms showing the enhanced ``magnetic transientï'' features
associated with the flare during the same period as above along the
dotted and dashed curves L1 and L2.\label{fl:two}}
\end{figure}

The non-LTE calculations for Ni {\sc i} line have shown that this
absorption line can turn into emission only by a large increase of
electron density and not by thermal heating of the atmosphere by any
other means. Qiu \& Gary (2003), interestingly, noted that
sign-reversals occurred near locations of strong magnetic field,
cooler sunspot umbrae that are exactly co-aligned with HXR sources.
They suggested that such non-thermal effects are most pronounced in
cool atmosphere where continuum is maintained at low intensity
level. This implies that the sign reversal anomaly may be associated
with non-thermal electron beam precipitating to the lower atmosphere
near cool sunspots.

Photospheric transients as reported previously by MA09 were observed
recently by SDO-HMI during the first X-class flare of the current
solar cycle 24 on 2011 February 15 in AR NOAA 11158 as shown in
Figure 2. (Detailed study of this event is reported by Maurya,
Vemareddy \& Ambastha 2012). Unlike the Ni {\sc i} line used in MDI
and GONG, the SDO-HMI uses Fe {\sc i} 6173.3 \AA. It observes the
Sun in two circular polarizations at six wavelength positions
($\pm34.4, \pm103.2$ and $\pm172.0$\,m\AA) of the spectral line.
Using these observations one can construct the line profile for any
desired location of the full disk Sun (Martinez Oliveros et al.,
2011). Therefore, it is now possible to analyze the line profile
changes, if any, associated with energetic transients and its effects on the estimation of
magnetic and Doppler velocity fields.

\begin{figure}
\centerline{\includegraphics[width=11cm]{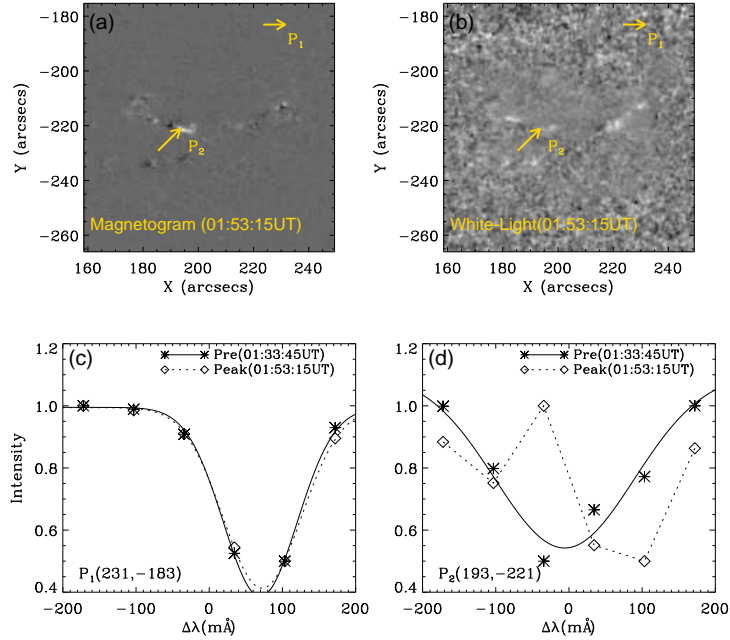}}
\caption{Top: (a) Consecutive difference magnetogram, and (b)
white-light images of the AR NOAA 11158 during peak phase (01:53:15
UT) of the X2.2 flare. Bottom: Line profiles at (c) a quiet location
P1, and (d) a transient location P2 as marked in top panel. Fe {\sc
I} line profiles are plotted during pre- (01:33:45 UT) and peak-
(01:53:15 UT) phases of the flare.\label{f:three}}
\end{figure}

We found spatial association of the transients with the WLF kernels
and HXR sources at the different phases of evolution of the X2.2
flare. Figure~3 shows the line profiles at a quiet location P$_1$
and at a location P$_2$ of the transient feature. Solid (dashed)
curves represent the line profiles during the pre (peak) phase of
the flare. We found that the line profile turned to emission at the
wavelength $\Delta\lambda$=--34.4\,m\AA~and became broader during
the peak phase of the flare as compared to the pre-flare phase.
 No such change was found at the quiet, reference location P$_1$.
For a quantitative study of the transient related changes in  the
Stokes profiles, we further analyzed the line profiles at several
flaring and quiet locations of the AR NOAA 11158. The Stokes profiles
{\it I}$\pm${\it U} and {\it I}$\pm${\it Q} at the transient
location P2 were also examined in order to ascertain the results on
the line shape changes during the large flare. We found that the
magnetic flux and Doppler velocity values as estimated by HMI
algorithms were affected by these spectral changes, resulting in the
observed transient features.

\section{Summary and Conclusions}\label{sec:concl}

It is evident from the spectral data available from SDO-HMI that the
Fe {\sc i} line profile changes occurred during the impulsive phase
of the energetic X2.2 flare  of 2011 February 15 in AR NOAA 11158.
The observed changes could be related to various physical processes
occurring during the flare as follows.

The thermodynamic change during the impulsive phase of the flare can
drastically change the height of line formation  due to the moderate
perturbation of temperature and density. This can also occur as a
result of the large increase of electron density as shown earlier by
the non-LTE calculations of Ding et al. (2002) for Ni {\sc i} 6768 \AA~ line.
They suggested that the non-thermal excitation and ionization by the
penetrating electrons generate a higher electron density which
enhances the continuum opacity, thereby pushing the formation height
of the line upward. The precipitation of electrons and deposition of
energy in the chromosphere, enhanced radiation in the hydrogen
Paschen continuum gives rise to the line source function, leading to
an increase of the line core emission relative to the far wing and
continuum.

We thus conclude that the anomalous magnetic and Doppler transients
observed during the X2.2 flare of 2011 February 15 were related to
the Fe {\sc i} line profile changes. Both thermal and non-thermal
physical processes operating during the flare may contribute to the
spectral profile changes at different stages of the flare evolution.
Therefore, the observed magnetic and velocity transients are
essentially the observational signatures of these physical
processes operating during the energetic event, and do not correspond
to ``real'' changes in the magnetic (doppler) fields.

%------------------------------------------------------------------------------%
%\section*{Acknowledgements}

\label{lastpage}
%------------------------------------------------------------------------------%

\begin{thebibliography}{}
%
\bibitem{amb:93} Ambastha, A., Hagyard, M. ~J., West, E. ~A., 1993, Solar Phys., 148,
277.

\bibitem{amb:03} Ambastha, A., Basu, S., Antia, H.~M., 2003, Solar Phys. 218, 151.

\bibitem{bru:93} Bruls, J.~H.~M.~J., 1993, A \& A, 269, 509.

\bibitem{din:02} Ding, M. ~D., Qiu, ~J., Wang, ~H., 2002, ApJ, 576, L83.

\bibitem{har:86} Harvey, J., 1986, in Deinzer, W., Knolker, M., Voigt, H.~H., eds.,
Small Scale Magnetic Flux Concentrations in the Solar Photosphere,
Vandenhoeck \& Ruprecht, Goettingen, p.~25.

\bibitem{kos:11} Kosovichev, A. ~G., 2011, ApJ, 734, L15

\bibitem{kos:01} Kosovichev, A. ~G., Zharkova, V. ~V., 2001, ApJ, 550, L105

\bibitem{mar:11} Martinez Oliveros, J. ~C., Couvidat, S., Schou, J., et al., 2011,
Solar Phys., 269, 269

\bibitem{mat:00} Mathew, S.~K., Ambastha, A., 2000, Solar Phys. 197, 75.

%\bibitem{mau:08} Maurya, R.~A., Ambastha, A., 2008, J. Astrophys. Astron. 29, 103.

\bibitem{mau:09} Maurya, R. ~A., Ambastha, A., 2009, Solar Phys., 258, 31

\bibitem{mau:12} Maurya, R. ~A., Vemareddy P. , Ambastha, A., 2012, ApJ, 747:134
(doi: http://dx.doi.org/10.1088/0004-637X/747/2/134 )

\bibitem{patt:84} Patterson, A., 1984, ApJ, 280, 884

\bibitem{qiu:03} Qiu, J., Gary, D.~ E., 2003, ApJ, 599, 615

\bibitem{sud:05} Sudol, J.~J., Harvey, J.~W., 2005, ApJ,  635, 647.

\bibitem{wan:02} Wang, H. et al. 2002, ApJ, 576, 497.

%\bibitem{wol:72} Wolff, C.~L., 1972, ApJ, 176, 833.

\bibitem{zha:07} Zharkova, V. ~V., Zharkov, S. ~I., 2007, ApJ, 664, 573

\bibitem{zir:81} Zirin, H., Tanaka, K., 1981, ApJ, 250, 791.
%
%
\end{thebibliography}
\end{document}